\documentclass[
twocolumn,
showpacs,preprintnumbers,
bibnotes,
 amsmath,amssymb,
 aps,
 pra,
superscriptaddress,
longbibliography,
]{revtex4-1}

\usepackage[version=3]{mhchem} 
\usepackage[dvipsnames]{xcolor}
\usepackage{textcomp}
\usepackage{graphicx}
\usepackage{amsmath}
\usepackage{fixmath}
\usepackage{amsfonts}
\usepackage{amssymb}

\usepackage[colorlinks=true,citecolor=blue]{hyperref}
\hypersetup{colorlinks=true,citecolor=blue,linkcolor=blue,urlcolor=blue}

\graphicspath{ {./figures/} }

\begin{document}

\title{Electrostatic control of nonlinear photonic-crystal polaritons in a monolayer semiconductor}

\author{Ekaterina Khestanova}
\affiliation{School of Physics and Engineering, ITMO University, Saint Petersburg 197101, Russia}

\author{Vanik Shahnazaryan}
\affiliation{School of Physics and Engineering, ITMO University, Saint Petersburg 197101, Russia}

\author{Valerii K. Kozin}
\affiliation{Department of Physics, University of Basel, Klingelbergstrasse 82, CH-4056 Basel, Switzerland}

\author{Valeriy I. Kondratyev}
\affiliation{School of Physics and Engineering, ITMO University, Saint Petersburg 197101, Russia}

\author{Dmitry N. Krizhanovskii}
\author{Maurice S. Skolnick}
\affiliation
{Department of Physics and Astronomy, University of Sheffield, Sheffield S3 7RH, UK}

\author{Ivan A. Shelykh}
\affiliation{Science Institute, University of Iceland, Dunhagi 3, IS-107, Reykjavik, Iceland}
\affiliation{School of Physics and Engineering, ITMO University, Saint Petersburg 197101, Russia}

\author{Ivan V. Iorsh}
\author{Vasily Kravtsov}
\email{vasily.kravtsov@metalab.ifmo.ru}
\affiliation{School of Physics and Engineering, ITMO University, Saint Petersburg 197101, Russia}

\date{\today}


\begin{abstract}
\noindent 
\textbf{
Integration of 2D semiconductors with photonic crystal slabs provides an attractive approach to achieve strong light--matter coupling and exciton-polariton formation in a planar chip-compatible geometry.
However, for the development of practical devices, it is crucial that the polariton excitations in the structure are easily tunable and exhibit strong nonlinear response.
Here we study neutral and charged exciton-polaritons in an electrostatically gated planar photonic crystal slab with an embedded monolayer semiconductor MoSe$_2$ and experimentally demonstrate strong polariton nonlinearity, which can be tuned via gate voltage.
We find that modulation of dielectric environment within the photonic crystal results in the formation of two distinct resonances with significantly different nonlinear response, which enables optical switching with ultrashort laser pulses.
Our results open new avenues towards development of active polaritonic devices in a compact chip-compatible implementation. 
}
\end{abstract}

\maketitle

\noindent
Integrated nanophotonics provides a potential route towards faster and more energy-efficient processing of classical and quantum information in smaller volumes~\cite{Koenderink2015}.
However, the development of active integrated nanophotonic circuits~\cite{Cheben2018} requires optical elements with high nonlinearity, tunability, and chip compatibility, which are not readily provided by traditional optical materials and structures.

A promising approach to enhance optical nonlinearity is enabled by strong light--matter coupling and formation of, for example, exciton-polaritons in semiconductor microcavities~\cite{Skolnick1998}.
These quasiparticles, while partially retaining their photonic character, can acquire a strongly nonlinear behavior~\cite{Khitrova1999} through the significant exciton--exciton interaction.
Recently, strongly coupled systems based on atomically thin transition metal dichalcogenides (TMDs) have emerged as a viable platform for polaritonics~\cite{Schneider2018,Hu2020} as they are characterized by large Rabi splittings, high exciton binding energies, relative ease of fabrication, and sizeable nonlinear interaction~\cite{Barachati2018,Gu2021,Zhang2021}.

The two-dimensional nature of atomically thin TMDs allows straightforward tuning of their free carrier density via electrostatic gating~\cite{Beck2020}, which enables active control of both linear~\cite{Ross2013} and nonlinear~\cite{Seyler2015,shahnazaryan2020tunable} optical properties.
This leads to a high degree of tunability for exciton-polaritons in TMD-based structures via free-carrier-induced modification of the 2D exciton oscillator strength and Rabi splitting, which has been demonstrated in various microcavity-based~\cite{Sidler2017,Flatten2017,Chakraborty2018} and plasmonic~\cite{Lee2017,Dibos2019,Munkhbat2020} systems with embedded TMD monolayers.
Beyond the control on neutral 2D exciton-polaritons, electrostatic doping enables the formation of novel charged polariton states that exhibit enhanced nonlinear interactions~\cite{Emmanuele2020,Tan2020,Kyriienko2020,lyons2022giant}.
However, so far the electrostatic tunability of interacting TMD exciton-polaritons have been studied in structures poorly suited for chip integration.

Recently, TMD polaritons have been demonstrated in chip-compatible systems that take advantage of dielectric photonic crystal slabs~\cite{Zhang2018,Gogna2019,Kravtsov2020,Chen2020} and thus eliminate the shortcomings associated with the use of bulky Bragg cavities or lossy plasmonic resonators.
The results of first experiments on nonlinear~\cite{Kravtsov2020} and charged~\cite{Koksal2021} photonic-crystal TMD polaritons call for further investigation of the electrostatic control on polaritons and their nonlinear response in these technologically important hybrid light--matter systems.

Here we experimentally study exciton-polaritons in an electrostatically gated MoSe$_2$ monolayer integrated with a planar hybrid Ta$_2$O$_5$/hBN dielectric photonic crystal.
We show that periodic modulation of the dielectric environment leads to the formation of two distinct exciton states that co-exist within a unit cell of the photonic crystal, with two additional trion states formed at finite free carrier density.
The exciton states are strongly coupled to the optical fields, forming polaritons that are highly tunable via electrostatic gate voltage.
We study the nonlinear response of the resulting exciton- and trion-polariton resonances and show that it can be controlled both electrostatically and optically.
Our results demonstrate that integration of photonic crystal slabs with gated monolayer semiconductors provides a highly tunable, nonlinear, and chip-compatible strongly coupled system for future development of polariton-based compact active optoelectronic devices.



\begin{figure*}[t]
	\includegraphics[width=\textwidth]{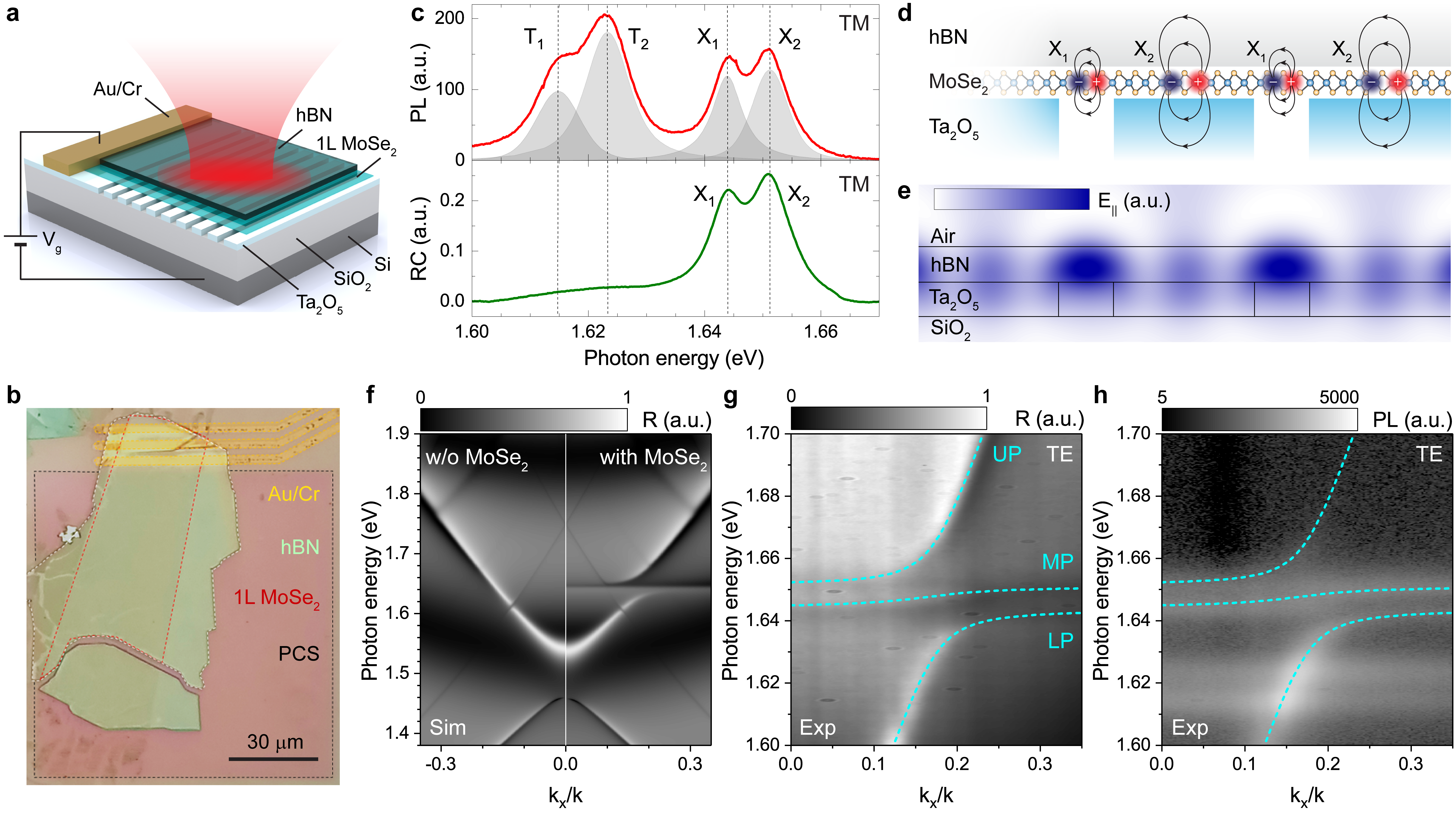}
	\caption{
 \textbf{Photonic-crystal exciton-polaritons at zero gate voltage $V_g = 0$~V. a} Schematic of the studied device consisting of Ta$_2$O$_5$ photonic crystal slab on SiO$_2/$Si substrate integrated with MoSe$_2$ monolayer and covered with multilayer hBN, with Au/Cr electrodes for applying gate voltage between the monolayer and Si. \textbf{b} Optical microscope image of the fabricated sample. \textbf{c} Photoluminescence (top plot) and reflectivity contrast (bottom plot) spectra measured in the TM polarization, with indicated exciton (X$_1$, X$_2$) and trion (T$_1$, T$_2$) resonances. \textbf{d} Schematic illustration of the 2 spatially localized exciton species formed in MoSe$_2$ due to periodic modulation of the dielectric environment. \textbf{e} Calculated spatial distribution of the in-plane E-field amplitude in the studied structure under plane-wave excitation at 10$^{\circ}$ incidence. \textbf{f} Simulated angle-resolved reflection spectra for the structure without (left) and with (right) monolayer MoSe$_2$. \textbf{g} Angle-resolved reflection spectra measured in the TE polarization. \textbf{h} Angle-resolved photoluminescence spectra measured in the TE polarization. Dashed curves are fits according to the model of 3 coupled oscillators.
 }
	\label{fig:Device}
\end{figure*}

\begin{figure*}[t]
	\includegraphics[width=\textwidth]{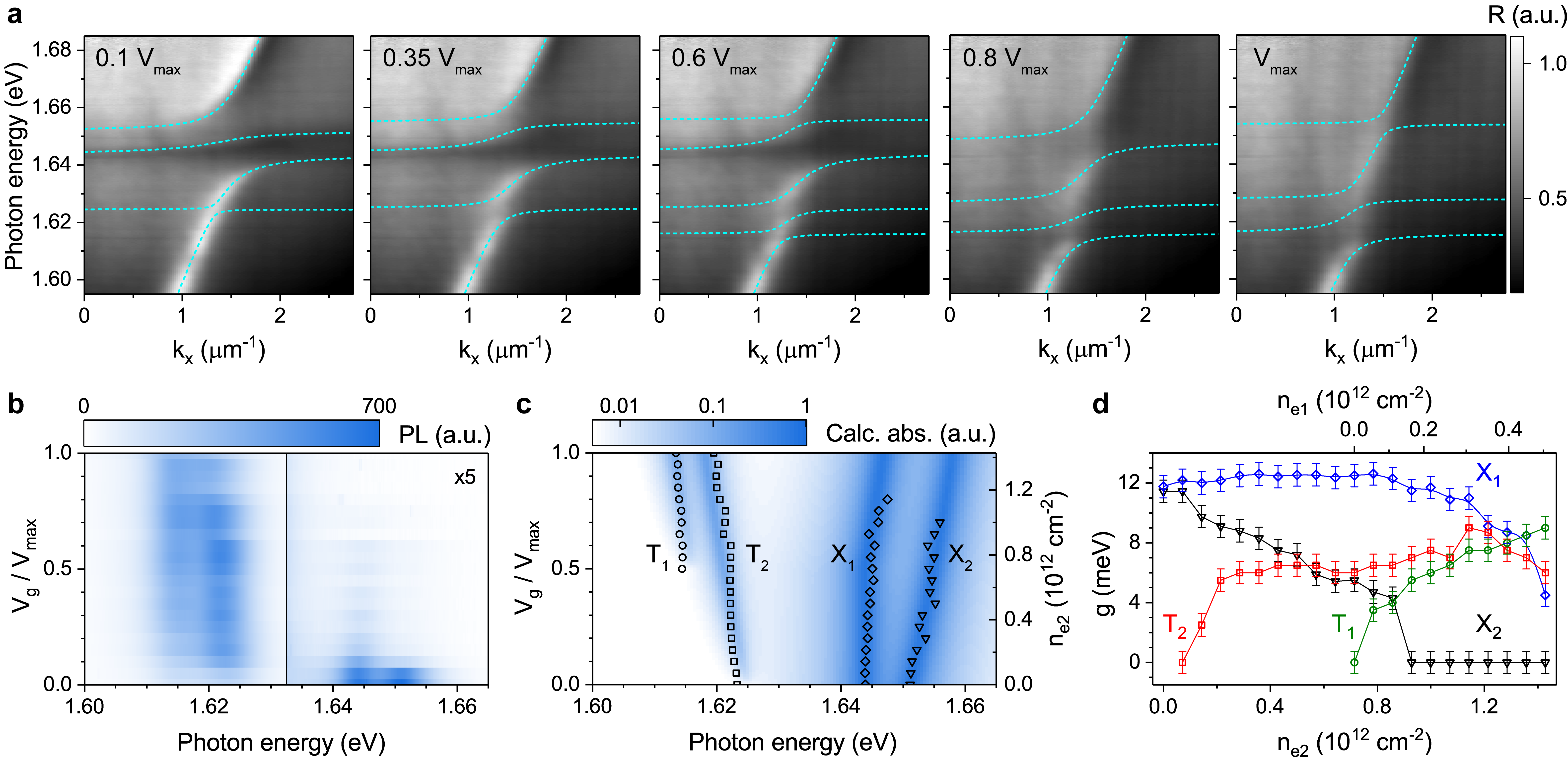}
	\caption{
 \textbf{Electrostatic control of the exciton- and trion-polaritons in the linear regime. a} Angle-resolved reflection spectra measured in the TE polarization at different gate voltages indicated in each panel. Dashed curves are fits according to the coupled oscillators model. \textbf{b} Photoluminescence spectra measured in the TM polarization for variable gate voltage. \textbf{c} Calculated absorption spectra for variable density of free electrons (false color plot) together with extracted from experimental data peak positions of the 2 trion (circles and squares) and 2 exciton (diamonds and triangles) resonances. \textbf{d} Coupling strength values for excitons X$_1$ and X$_2$ (blue diamonds and black triangles) and trions T$_1$ and T$_2$ (green circles and red squares) extracted from coupled oscillator model fits to the experimental polariton dispersions.
 }
	\label{fig:Gating}
\end{figure*}

Our experimental sample for investigating electrostatically tunable nonlinear polaritons consists of a MoSe$_2$ monolayer sandwiched between two high-index dielectric slabs of sub-100~nm thickness: Ta$_2$O$_5$ on bottom and multilayer hexagonal boron nitride (hBN) on top, as schematically illustrated in Fig.~\ref{fig:Device}a.
The Ta$_2$O$_5$ slab is periodically modulated by etched grooves, forming a 1D photonic crystal on top of a Si substrate with a thick (1 um) SiO$_2$ oxide layer.
The MoSe$_2$ monolayer is in contact with metal electrodes deposited on the Ta$_2$O$_5$ surface, which allows us to control the free carrier density in MoSe$_2$ by applying a gate voltage between the electrode and Si substrate.
An optical microscope image of the sample is shown in Fig.~\ref{fig:Device}b (see Methods for details on fabrication).

We probe uncoupled exciton states in the studied system via reflectivity contrast (RC) and photoluminescence (PL) measurements in the TM polarization (E-field perpendicular to grooves) averaged over a large sample area.
Experimental spectra for zero gate voltage $V_g = 0$~V shown in Fig.~\ref{fig:Device}c reveal 2 neutral exciton resonances X$_1$ and X$_2$ at $\omega_\mathrm{X1}\simeq 1.644$~eV and $\omega_\mathrm{X2}\simeq 1.651$~eV for both PL (top) and RC (bottom) signals, with PL spectra showing 2 additional lower-energy charged exciton (trion) peaks T$_1$ and T$_2$ caused by substrate-induced residual doping of the MoSe$_2$ monolayer.
We attribute the observed double-peaked spectra to the formation of 2 distinct spatially localized excitonic states resulting from the periodic spatial modulation of the refractive index in the  Ta$_2$O$_5$ slab, which locally modifies the dielectric environment for MoSe$_2$ (see also results of spatial mapping in Supplementary Figure~1).
As schematically illustrated in Fig.~\ref{fig:Device}d, excitons X$_1$ are created in the groove regions where the MoSe$_2$ monolayer is interfaced only with the hBN surface, while excitons X$_2$ are created in the ridge regions where MoSe$_2$ is sandwiched between hBN and Ta$_2$O$_5$.
Due to the high dielectric constant of Ta$_2$O$_5$ at optical frequencies of $\epsilon_\mathrm{Ta_2O_5}\sim 4.4$, screening in the MoSe$_2$ monolayer can be significantly modified when interfaced with Ta$_2$O$_5$, resulting in a spectral blue shift of the X$_2$ resonance frequency in comparison to X$_1$.
We note that possibly different local strain in the groove and ridge spatial regions might additionally contribute toward the formation of two spectrally separated excitonic states.

The two excitonic resonances are expected to strongly couple with local optical fields in the photonic crystal structure.
Fig.~\ref{fig:Device}e shows the calculated spatial distribution for the in-plane component of the local electric field under plane-wave excitation at 10$^{\circ}$ with the normal.
While E-field is maximized inside the hBN layer above grooves, the overlap with the MoSe$_2$ monolayer is stronger in the ridge areas, resulting in efficient exciton--photon coupling in both spatial regions.
The effect of exciton--photon coupling on the reflectivity of the structure in the TE polarization (E-field parallel to grooves) is illustrated in Fig.~\ref{fig:Device}f, where we plot results of numerical simulations performed using the Fourier modal method (FMM, see also Supplementary Figure~2).
The dispersion of the hBN/Ta$_2$O$_5$ photonic crystal is shown in the left panel, while the dispersion of the full structure including monolayer MoSe$_2$ is shown in the right panel and exhibits a clear anticrossing between the optical mode and exciton resonance around 1.65~eV forming distinct exciton-polariton branches. 

We probe the resulting polaritons experimentally using angle-resolved reflection and PL spectroscopic measurements (see Methods for details).
The reflection and PL data measured at zero gate voltage in the TE polarization are shown in Fig.~\ref{fig:Device}g and Fig.~\ref{fig:Device}h, respectively, and demonstrate a clear anticrossing between the photonic crystal mode and X$_1$, X$_2$ excitonic resonances, characteristic of the strong light--matter coupling regime.
As a result, 3 polariton branches are formed in the dispersion corresponding to the lower (LP), middle (MP), and upper (UP) polariton states, as indicated in Fig.~\ref{fig:Device}g,h with dashed cyan curves obtained by fitting the experimental data with a model of 3 coupled oscillators.

We control the polariton dispersion electrostatically by varying the gate voltage $V_g$ as shown in Fig.~\ref{fig:Gating}a with experimental angle-resolved reflection spectra for selected values of $V_g$ as a fraction of the maximum applied voltage of $V_\mathrm{max} = 500$~V.
At small non-zero $V_g = 0.1 V_\mathrm{max}$ (left panel), the dispersion starts to exhibit splitting on the T$_2$ trion resonance at $\omega_\mathrm{T2} = 1.623$~eV.
At $V_g = 0.5 V_\mathrm{max}$ (middle panel), additional splitting emerges on the T$_1$ trion resonance at $\omega_\mathrm{T1} = 1.614$~eV.
With increasing gate voltage, both splittings get stronger and at $V_g = V_\mathrm{max}$ dominate the measured polariton dispersion, whereas the splitting on the neutral exciton states X$_{1,2}$ is gradually reduced.
Reflection spectra for selected wave vector values $k_x$, corresponding to the maps shown in Fig.~\ref{fig:Gating}a, are plotted in Supplementary Figure~3.

We fit the experimental data with a model of 5 coupled oscillators representing exciton resonances X$_1$ and X$_2$, trion resonances T$_1$ and T$_2$, and optical mode in the photonic crystal structure.
The spectral positions of the exciton and trion resonances are taken from gate voltage dependent PL and RC measurements of uncoupled states in the TM polarization shown in Fig.~\ref{fig:Gating}b and Supplementary Figure~4, respectively.
The extracted peak positions are plotted in Fig.~\ref{fig:Gating}c as symbols and demonstrate good agreement with theoretically modelled absorption spectra shown in the false color plot (see Supplementary Note~1 for calculation details).
With increasing gate voltage, both exciton resonances X$_{1, 2}$ shift to higher energies, while trion resonances T$_{1, 2}$ exhibit a slight shift toward lower energies.

From fits of the experimentally measured polariton dispersions, which are plotted in Fig.~\ref{fig:Gating}a with dashed cyan curves, we extract the values of coupling strength $g_i$ for the two exciton and two trion resonances.
The obtained values are shown in Fig.~\ref{fig:Gating}d as functions of free carrier density $n_e$ controlled by the applied gate voltage $V_g$. 
We note that, for the same $V_g$, free electron densities in the groove and ridge regions of the photonic crystal ($n_{e1}$ and $n_{e2}$) are different due to the different capacitance and residual doping.
We convert gate voltage to $n_{e1}$ and $n_{e2}$ values as outlined in Supplementary Note~2.
With increasing electron density, the coupling strength values for both trion resonances (green circles and red squares) grow approximately as $\sqrt{n_e}$, which we attribute to the increase in the corresponding trion oscillator strength values \cite{Zhumagulov2022}.
At high electron densities, a significant portion of neutral excitons bind to electrons to form trions, and the exciton oscillator strength starts to decrease \cite{shahnazaryan2020tunable}, which is observed in our data as the decreasing coupling strength for both neutral exciton resonances (blue rhombuses and black triangles).

\begin{figure*}[t]
	\includegraphics[width=\textwidth]{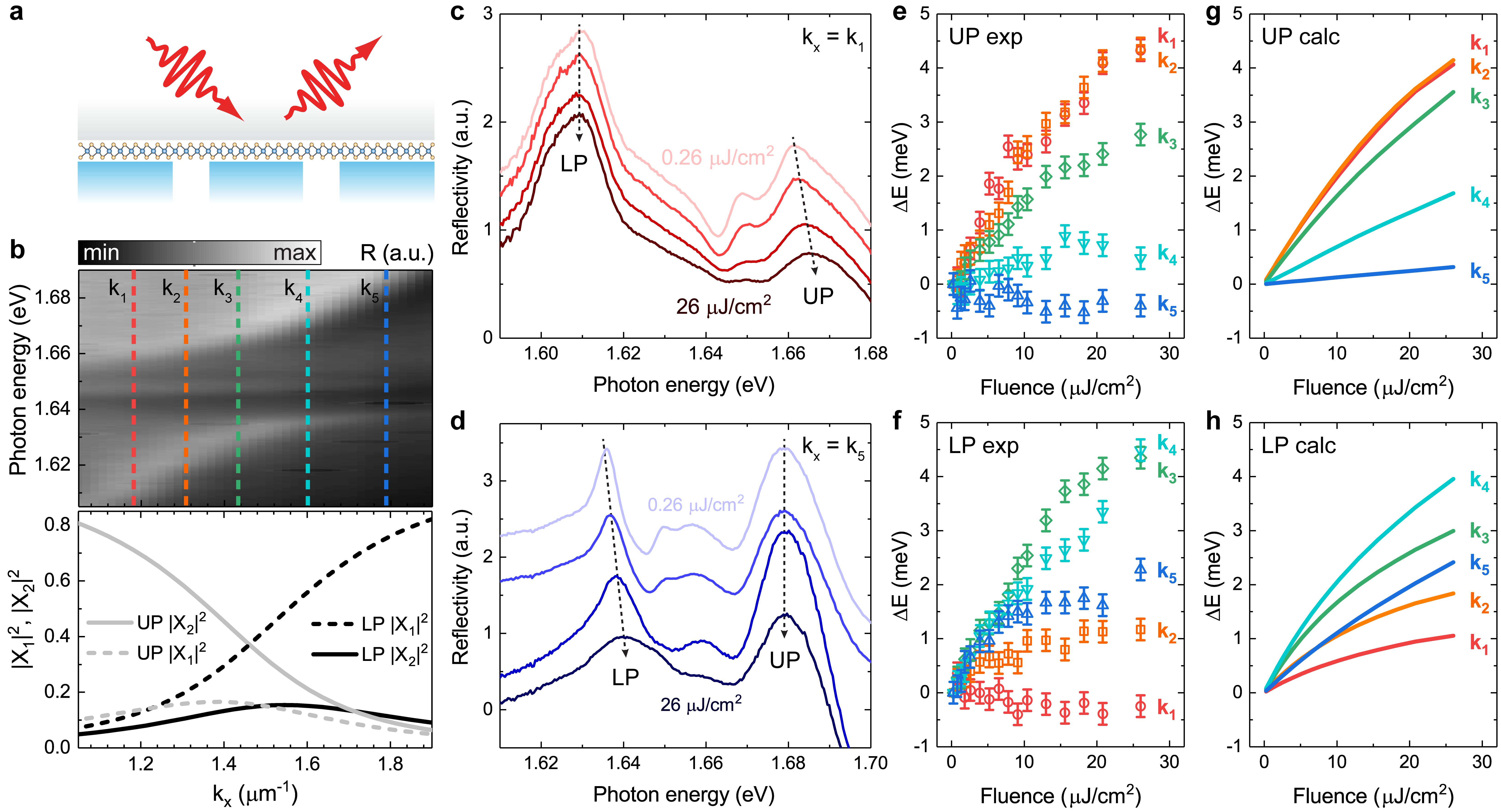}
	\caption{
 \textbf{Resonant nonlinear response of photonic-crystal exciton-polaritons at zero gate voltage. a} Measurement schematic showing femtosecond laser pulses that excite polaritons resonantly in frequency and wavevector and are detected in the reflection geometry.
 \textbf{b} Linear dispersion of exciton-polaritons at $V_g = 0$~V with indicated values of 5 different excitation in-plane wavevectors (top panel) and corresponding calculated Hopfield coefficients showing fractions of excitons X$_1$ and X$_2$ for the upper and lower polariton branches.
 \textbf{c} Reflectivity spectra measured at low in-plane wavevector $k_1 = 1.182~\mu$m$^{-1}$ for excitation fluence values increasing from $F_\mathrm{min} = 0.26$~$\mu$J/cm$^2$ for the top curve to $F_\mathrm{max} = 26$~$\mu$J/cm$^2$ for the bottom curve.
 \textbf{d} Corresponding reflectivity spectra measured at high in-plane wavevector $k_1 = 1.789~\mu$m$^{-1}$.
 \textbf{e} Experimental fluence-dependent spectral shifts extracted for the UP branch at 5 different in-plane wavevectors.
 \textbf{f} Corresponding spectral shifts extracted for the LP branch.
 \textbf{g} Calculated fluence dependencies of the UP branch spectral shifts at the 5 in-plane wavevector values.
 \textbf{h} Corresponding fluence dependencies of the LP branch spectral shifts.
 }
	\label{fig:Power}
\end{figure*}

The demonstrated electrostatic tuning of the polariton dispersion further allows us to control polariton--polariton interaction and the associated optical nonlinearity.
To probe the nonlinear polaritonic response in our device, we excite it with 60~fs laser pulses resonantly in frequency and wavevector and measure the RC spectrum as a function of excitation fluence as schematically illustrated in Fig.~\ref{fig:Power}a.

We first discuss results obtained at zero gate voltage where we have only neutral exciton polaritons forming 3 branches in the dispersion corresponding to the lower polariton (LP), middle polariton (MP), and upper polariton (UP) states.
The dispersion measured in the linear optical regime is shown for reference in Fig.~\ref{fig:Power}b (top panel).
Polariton nonlinearity stems from exciton--exciton interaction and is therefore expected to depend sensitively on the exciton fraction in the polariton quantified by the Hopfield coefficient $|X|^2$.
We probe the nonlinear response under excitation at 5 different values of the in-plane wavevector component $k_x$ indicated in the figure with vertical dashed lines, which correspond to different Hopfield coefficients plotted in the bottom panel as functions of the in-plane wavevector $k_x$.
We analyze only the LP and UP resonances as the most pronounced in the experimentally measured spectra.
For each polariton branch, there are 2 excitonic Hopfield coefficients  $|X_1|^2$ and $|X_2|^2$ describing the fractions of the two excitons X$_{1,2}$ in the polariton.
We note that, according to the plot, the dominant contributions to the nonlinear behavior of the LP and UP resonances are provided by the X$_1$ and X$_2$ exciton states, respectively.

Experimentally measured reflectivity spectra at the lowest excitation in-plane wavevector $k_1 = 1.182~\mu$m$^{-1}$ are shown in Fig.~\ref{fig:Power}c, with excitation fluence increasing from $F_\mathrm{min} = 0.26$~$\mu$J/cm$^2$ for the top curve to $F_\mathrm{max} = 26$~$\mu$J/cm$^2$ for the bottom curve.
The presented spectra exhibit a distinct blueshift of the higher energy UP peak with increasing fluence, while the lower energy LP peak position remains unchanged.
Analogous reflectivity spectra measured at the highest excitation in-plane wavevector $k_5 = 1.789~\mu$m$^{-1}$ are shown in Fig.~\ref{fig:Power}d and reveal, in contrast, a distinct blueshift of the LP peak with increasing fluence, while the UP peak position remains unchanged.
From the obtained experimental spectra, we extract the LP and UP peak positions as functions of excitation fluence and plot them in Fig.~\ref{fig:Power}e and Fig.~\ref{fig:Power}f, respectively, where different colors correspond to different excitation wavevectors.
Within the experimental uncertainty, all peaks exhibit blueshift with increasing fluence, together with saturating behavior at higher fluence values.
For the UP branch (e), the blueshift is strongest at the lowest excitation wavevector $k_1$ (red circles), decreases for higher wavevectors (orange, green, and cyan symbols), and vanishes at $k_5$ (blue pyramids).
Conversely, for the LP branch (f), the blueshift generally increases together with excitation wavevector.
However, we note that the blueshift is decreased at $k_5$ (blue pyramids), which can be attributed to the less efficient excitation of the polariton resonance due to the effect of polariton motional narrowing significantly reducing the linewidth at higher wavevectors~\cite{Whittaker1996, Kondratyev2023}.

We model the nonlinear response of exciton-polaritons in our device and the associated pump power dependent spectral shifts via numerically solving coupled nonlinear dynamical equations for the two exciton and one optical fields as described in the Supplementary Note~3.
The results of modelling for the UP and LP branches are shown in Fig.~\ref{fig:Power}g and Fig.~\ref{fig:Power}h, respectively, and reproduce the experimentally observed trends.
Our model generally takes into account 2 mechanisms of nonlinearity including exciton exchange interaction and quench of Rabi splitting due to phase space filling effects.
However, best agreement with the experimental data, where both UP and LP branches exhibit blueshifts, is achieved when we consider the dominant contribution from exciton exchange interaction, which suggests that the quenching of Rabi splitting is small at the range of experimental excitation fluence values.
The corresponding exciton--exciton interaction strength is obtained as $V_\mathrm{XX1} = 1.066$ $\mu\mathrm{eV}\cdot\mu \mathrm{m}^2$ and $V_\mathrm{XX2} = 1.068$ $\mu\mathrm{eV}\cdot\mu \mathrm{m}^2$ for the groove and ridge regions in the photonic crystal unit cell, respectively.
These values are in general agreement with previous results for similar samples~\cite{Kravtsov2020,shahnazaryan2020tunable,shahnazaryan2017exciton,erkensten2021exciton,cam2022symmetry,bleu2020polariton,fey2020theory,Benimetskiy2023}. 
We note that the saturation of blueshifts at high fluence values is primarily due to exciton--exciton annihilation leading to a sub-linear dependence of polariton density on pump fluence. 
An additional mechanism limiting the range of nonlinear spectral shifts detected in the experiment is related to three-body excitonic interactions taken into account in the theoretical model.

\begin{figure*}[t]
	\includegraphics[width=\textwidth]{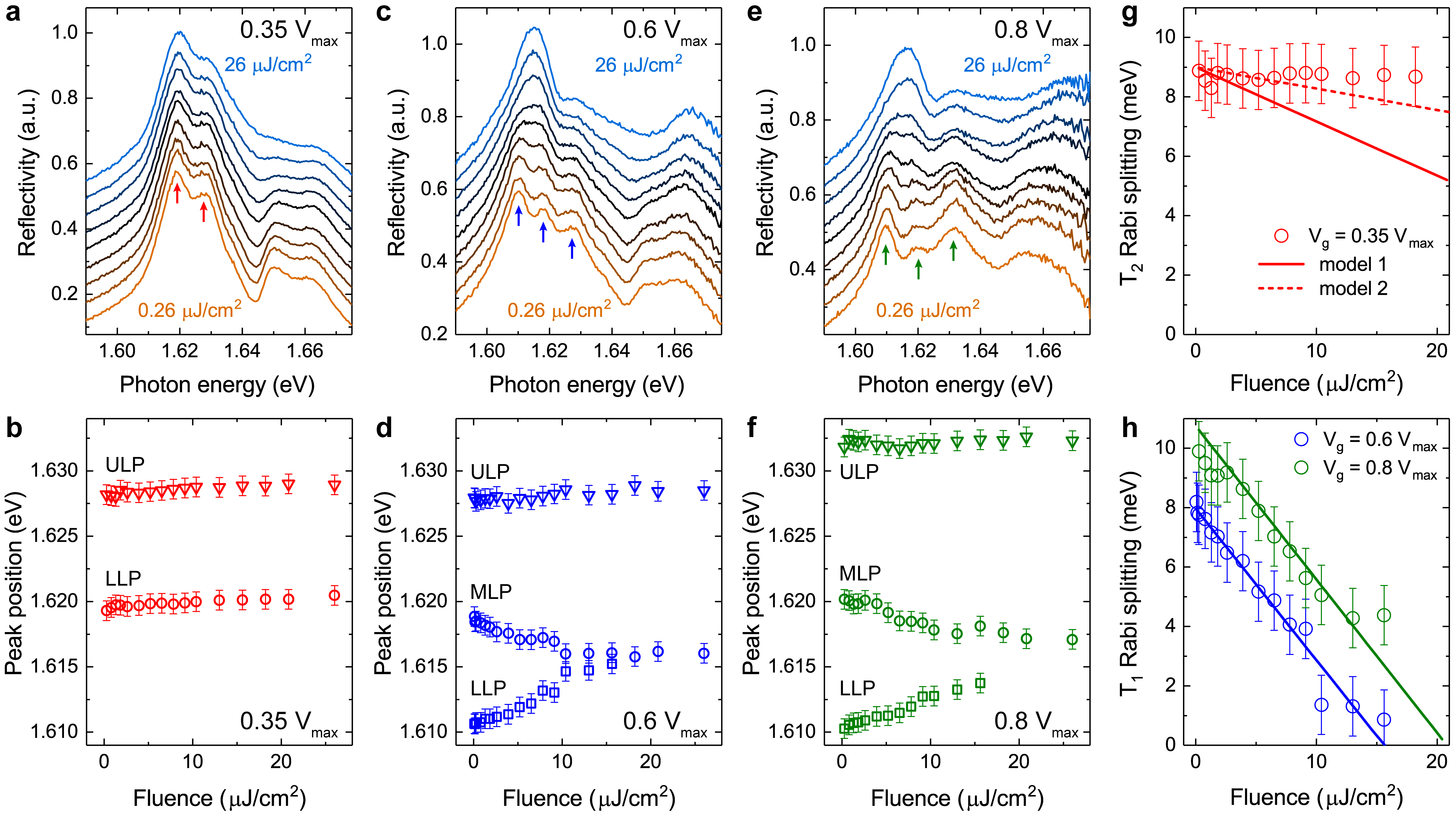}
	\caption{
 \textbf{Electrostatic tuning of nonlinear spectral shifts. a} Experimental reflectivity spectra measured in the TE polarization on the T$_2$ resonance for $V_g = 0.35 V_\mathrm{max}$ and excitation fluence varying from $F_\mathrm{min} = 0.26$~$\mu$J/cm$^2$ for the bottom yellow curve to $F_\mathrm{max} = 26$~$\mu$J/cm$^2$ for the top blue curve.
 \textbf{b} Corresponding extracted peak positions for the LLP (circles) and ULP (triangles) resonances indicated in (a) with red arrows.
 \textbf{c} Fluence-dependent experimental reflectivity spectra measured on the T$_1$ resonance for $V_g = 0.6 V_\mathrm{max}$.
 \textbf{d} Corresponding extracted peak positions for the LLP (squares), MLP (circles), and ULP (triangles) resonances indicated in (c) with blue arrows.
 \textbf{e} Fluence-dependent experimental reflectivity spectra measured on the T$_1$ resonance for $V_g = 0.8 V_\mathrm{max}$.
 \textbf{f} Corresponding extracted peak positions for the LLP (squares), MLP (circles), and ULP (triangles) resonances indicated in (e) with green arrows.
 \textbf{g} Rabi splitting (circles) as a function of fluence for the T$_2$ resonance obtained as a difference between the ULP and LLP energies shown in (b). Theoretical estimated based on average and min-max fluence variation are plotted with solid and dashed lines, respectively.
 \textbf{h} Rabi splitting as a function of fluence for the T$_2$ resonance at $V_g = 0.6 V_\mathrm{max}$ (blue circles) and $V_g = 0.8 V_\mathrm{max}$ (green circles) obtained as difference between the MLP and LLP energies shown in (d) and (f), respectively. Corresponding modelling results are plotted with blue and green lines.
 }
	\label{fig:Trions}
\end{figure*}

Next, we discuss results of nonlinear measurements at different free electron densities controlled by gate voltage.
For the neutral exciton resonance, we find that the corresponding nonlinear polaritonic response does not exhibit any pronounced dependence on free electron density.
We probe the corresponding nonlinearity via fluence-dependent spectral shift of the UP resonance and present experimental results for a selected wavevector value and different gate voltage values in Supplementary Figure~5.
The data reveal blueshift of the UP resonance with increasing pump fluence, with the trend similar to that observed at $V_g = 0$ and almost independent of gate voltage up to $V_g = 0.8 V_\mathrm{max}$.
This behavior is consistent with our theoretical model that predicts only a slight increase of the exciton--exciton coupling strength $V_\mathrm{XX}$ of less than 10$\%$ for finite electron density within the range of gate voltages applied in the experiment (see Supplementary Figure~6).

At finite free electron densities corresponding to non-zero $V_g$, the LP branch splits on the trion resonances leading to the formation of trion-polariton states as we demonstrated in Fig.~\ref{fig:Gating}.
Fig.~\ref{fig:Trions} presents our measurement results for the nonlinear response of these trion-polariton states.
Power-dependent reflectivity spectra for $V_g = 0.35 V_\mathrm{max}$ and in-plane wavevector $k_x = 1.266~\mu$m$^{-1}$ corresponding to the T$_2$ trion resonance are plotted in Fig.~\ref{fig:Trions}a.
The split upper LP (ULP) and lower LP (LLP) polariton states appear as two peaks in the reflectivity spectrum indicated with red arrows in the plot.
As the excitation fluence increases from $F_\mathrm{min} = 0.26$~$\mu$J/cm$^2$ for the bottom orange curve to $F_\mathrm{max} = 26$~$\mu$J/cm$^2$ for the top blue curve, both ULP and LLP peaks exhibit slight blueshift and broadening, with the extracted spectral positions plotted in Fig.~\ref{fig:Trions}b.

At higher free electron density, an additional splitting of the LP branch develops at the T$_1$ trion resonance.
We plot power-dependent reflectivity spectra for $V_g = 0.6 V_\mathrm{max}$ and in-plane wavevector $k_x = 1.182~\mu$m$^{-1}$ corresponding to the T$_1$ trion resonance in Fig.~\ref{fig:Trions}c, where 3 split polariton peaks are indicated with blue arrows.
The extracted spectral positions of the ULP, middle LP (MLP), and LLP states as functions of excitation fluence are plotted in Fig.~\ref{fig:Trions}d.
With increasing fluence, the LLP and MLP peaks rapidly move towards each other indicating the collapse of the T$_1$ Rabi splitting.
Similar behavior is also observed for $V_g = 0.8 V_\mathrm{max}$, with power-dependent reflectivity spectra shown in Fig.~\ref{fig:Trions}e and extracted ULP, MLP, and LLP peak positions shown in Fig.~\ref{fig:Trions}f.
While the T$_1$ Rabi splitting is higher at this gate voltage, its power-dependent collapse occurs with a rate similar to that observed in (d).

The observed trends of the excitation-induced Rabi splitting collapse for the T$_2$ and T$_1$ trion resonances are summarized in Fig.~\ref{fig:Trions}g and h, respectively.
For the T$_1$ resonance (h), Rabi splitting exhibits a near-linear decrease with fluence.
We attribute this behavior to quenching of the T$_1$ trion oscillator strength for increasing trion density caused by phase space filling~\cite{Emmanuele2020,Kyriienko2020}.
Rabi splitting dependence on trion density $n_T$ in this case can be approximated as $\Omega_R(n_T) \simeq \Omega_R(0)\left(1 - 2n_{T}/n_{e}\right)$, where $n_{e}$ is the local free electron density, and $\Omega_R(0)$ is Rabi splitting in the linear regime $n_T \ll n_e$. 
Considering that trion density is proportional to fluence $n_T = \alpha F$, where excitation rate $\alpha$ scales with electron density $n_e$, trion Rabi splitting is expected to exhibit a linear decrease in fluence with a slope almost independent of $n_e$ (see Supplementary Note 4).
Theoretical dependencies $\Omega_R(F)$ calculated for two gate voltages $V_g = 0.6 V_\mathrm{max}$ and $V_g = 0.8 V_\mathrm{max}$ are plotted in Fig.~\ref{fig:Trions}h (blue and green lines, respectively) and show good agreement with experimental data (circles).

Surprisingly, the experimentally obtained Rabi splitting for the T$_2$ trion resonance exhibits a substantially weaker dependence on excitation fluence, as seen in Fig.~\ref{fig:Trions}g (red circles).
This can be attributed in part to a reduced trion excitation rate due to the lower local fluence.
As we show in Supplementary Figure 7, the spatial variation of in-plane local field across the photonic crystal unit cell results in the average fluence being $\sim 2.7$ times higher in the groove (vacuum/MoSe$_2$/hBN) than in the ridge (Ta$_2$O$_5$/MoSe$_2$/hBN) region.
Thus the slope of $\Omega_R(F)$ for the T$_2$ resonance is expected to be reduced compared to that shown in (h) for the T$_1$ resonance. 
The corresponding theoretical dependence is plotted in Fig.~\ref{fig:Trions}g with solid red line and labeled ``model 1''.
Further, non-uniform spatial distribution of free electron density within the photonic crystal cell can potentially affect the observed nonlinear response.
For example, sagging and associated strain potential in the groove region can cause funneling of the electrons and trion-polaritons towards the center of the groove where the local fluence is maximized.
Additionally, the trion-polariton response in the Ta$_2$O$_5$/MoSe$_2$/hBN region can be effectively generated at the minima of local fluence where phase space filling effects are weaker and more electrons are available for trion formation.
Accounting for these effects can further lower the slope of $\Omega_R(F)$ for the T$_2$ resonance as shown in Fig.~\ref{fig:Trions}g with dashed red line and labeled ``model 2''.

The observed different behavior of the nonlinear polariton response for the T$_1$ and T$_2$ trion resonances provides a new approach towards all-optical switching.
In the range of gate voltages where the LP branch is split on both the T$_1$ and T$_2$, polariton Rabi splitting for the T$_1$ resonance can be sensitively controlled by optical pump, with the possibility of full saturation of the T$_1$ resonance and effective switching to T$_2$ trion-polaritons.
We note that selective formation of T$_1$ and T$_2$ polariton states can be also achieved via electrostatic control, as we demonstrated in Fig.~\ref{fig:Gating}.
However, the potential speed of electrostatic switching is limited to the nanosecond scale by control electronics, device capacitance, and fundamental processes related to hole trapping at the interface between monolayer semiconductor and metal electrode~\cite{Zhu2023}.
In contrast, optical switching between T$_1$ and T$_2$ polariton states can be ultrafast.
In this work, we detect the nonlinear response via self-action of $\sim 60$~fs laser pulses, which implies that the associated nonlinearity is fast on the 10s of fs temporal scale.

In summary, we have investigated neutral and charged exciton-polaritons in an electrostatically tunable hybrid structure based on a photonic crystal slab with embedded monolayer semiconductor MoSe$_2$.
Spatial modulation of dielectric environment in the photonic crystal results in the formation of two distinct excitonic resonances in the monolayer, which are strongly coupled to the optical mode.
The resulting exciton- and trion-polaritons are tunable via gate voltage and exhibit efficient nonlinear response.
This enables optical switching between different polariton states using resonant excitation with femtosecond laser pulses.
Our results provide new opportunities for the development of active polaritonic devices in a compact chip-compatible implementation.\\

\noindent
{\bf \large Methods}\\
\noindent
\textbf{Sample fabrication.}
Ta$_2$O$_5$ layers of 90~nm thickness were deposited on commercial p-doped Si substrates with 1~$\mu$m dry thermal oxide via e-beam assisted ion-beam sputtering.
Photonic crystal slabs were fabricated by patterning the Ta$_2$O$_5$ layers via a combination of electron-beam lithography and ion beam etching.
The resulting gratings had the pitch of $p = 535$~nm, ridge width $r = 385$~nm, groove width $g = 150$~nm, and groove depth $d = 90$~nm, as measured with scanning electron microscopy and atomic force microscopy.
Metal contacts were fabricated on top of the unpatterned Ta$_2$O$_5$ layers near a photonic crystal slab (PCS) using electron-beam lithography, thermal evaporation, and lift-off.
The electrodes consisted of 20~nm thick Au on top of a 5~nm thick Cr adhesion layer.
Large-area high quality flakes of monolayer MoSe$_2$ and multilayer hBN were mechanically exfoliated from commercial bulk crystals (HQ Graphene) with Nitto tape.
The monolayer MoSe$_2$ flake was transferred onto the sample via dry transfer to simultaneously achieve optimal overlap with the PCS area and cover the electrodes.
Finally, the monolayer was capped with a 90~nm thick multilayer hBN flake to form the resulting structure.\\

\noindent
\textbf{Optical measurements.}
The sample was mounted in an ultra low vibration closed cycle helium cryostat (Advanced Research Systems) and maintained at a temperature of $7$~K.
The cryostat was mounted on a precise xyz stage for sample positioning.
The carrier density in the MoSe$_2$ monolayer was controlled via electrostatic gating by applying a gate voltage between the fabricated Au/Cr contacts and doped Si substrate with a sourcemeter Keithley 6487.
The leakage current was kept below 100~pA to avoid undesired sample heating effects.
Photoluminescence measurements were performed with off-resonant excitation by monochromatic light from a HeNe laser with wavelength $\lambda_\mathrm{exc} = 632.8$~nm.
Angle-resolved reflectance spectroscopy was performed in a back focal plane setup with a slit spectrometer coupled to a liquid nitrogen cooled imaging CCD camera (Princeton Instruments SP2500+PyLoN), using white light from a halogen lamp for illumination through a long working distance microscope objective (Mitutoyo 50X/0.42).
For pump-dependent reflectivity measurements, the sample was excited by 60~fs pulses from a wavelength-tunable femtosecond laser system (Light Conversion Orpheus optical parametric amplifier pumped by Pharos) with a repetition rate of 1~kHz.
Polaritons were excited and probed resonantly both in energy and wavevector.
The wavevector was controlled by focusing and positioning the laser beam within the back-focal plane of the microscope objective, yielding a Gaussian-shape laser spot on the sample with a full width at half maximum of $\sim 25~\mu$m.
Spatial filtering through a tunable rectangular slit was applied in the detection channel to selectively measure signals from the hBN/MoSe$_2$/PCS sample area.
Polarization was controlled with a thin-film polarizer in the detection channel.\\


\noindent
{\bf \large Acknowledgements}\\
\noindent
This research was supported by Priority 2030 Federal Academic Leadership Program.
The authors acknowledge funding from Ministry of Science and Higher Education of Russian Federation, goszadanie no. 2019-1246, and Russian Science Foundation project 21-72-10100.
V.S. acknowledges the support of ‘Basis’ Foundation (Project No. 22-1-3-43-1) and Goszadaniye No FSMG-2023-0011. V.K.K. acknowledges the support from the Georg H.
Endress Foundation.


%

\end{document}